\documentclass[prd,twocolumn,showpacs,floatfix,amsmath,nofootinbib,amssymb,floatfix]{revtex4}
\usepackage{graphicx,color,dcolumn,booktabs,bm}
\usepackage{longtable,lscape}
\usepackage{pdfpages}
\usepackage{txfonts}
\usepackage{overpic}
\usepackage{amssymb}
\usepackage{indentfirst}
\usepackage{feynmf}   
\usepackage{slashed}  
\usepackage{cases}
\usepackage{color}
\usepackage{multirow}
\usepackage{threeparttable}
\usepackage{epstopdf}
\usepackage{enumerate}
\usepackage{graphicx,color,dcolumn,booktabs,bm}
\usepackage[colorlinks, citecolor=blue,anchorcolor=red,menucolor=red, linkcolor=red,filecolor=red,urlcolor=blue,frenchlinks=red]{hyperref}

\graphicspath{{Figures/}} %

\begin{document}

\title{Assigning the newly reported $\Sigma_b(6097)$ as a $P$-wave excited state and predicting its partners}
\author{Bing Chen$^{1,3}$}\email{chenbing@shu.edu.cn}
\author{Xiang Liu$^{2,3}$\footnote{Corresponding author}}\email{xiangliu@lzu.edu.cn}
\affiliation{$^1$Department of Physics, Anyang Normal University,
Anyang 455000, China\\$^2$School of Physical Science and Technology,
Lanzhou University,
Lanzhou 730000, China\\
$^3$Research Center for Hadron and CSR Physics, Lanzhou University
$\&$ Institute of Modern Physics of CAS,
Lanzhou 730000, China}

\date{\today}

\begin{abstract}

The newly observed $\Sigma_b(6097)$ provides us a good chance to further construct the high excited states of the bottom baryon family. In this work, we explain the $\Sigma_b(6097)$ to be a $1P$ state with $J^P=5/2^-$ or $3/2^-$ by giving the mass spectrum analysis and the strong decay calculation, and predict the existence of $1P$ and $2S$ partners of the $\Sigma_b(6097)$, where their masses and the corresponding decay behaviors are presented. Because of the success of describing the $\Sigma_b(6097)$ and the former observed $\Xi_b(6227)$ under the quasi-two-body treatment to the bottom baryons, we continue to predict the $\Omega_b$ partners of $\Sigma_b(6097)$ by the same method, which are the $1P$ and $2S$ states of the $bss$ baryon system. Identifying these predicted bottom baryons will be a research topic full of challenge and opportunity for future experiments especially the LHCb.

\end{abstract}
 \maketitle

\section{Introduction}\label{sec1}
To accumulate the information of the hadron spectrum will help us to deepen the understanding of nonperturbative Quantum Chromodynamics (QCD). Among the whole hadron spectrum, the bottom baryon family, especially the excitations, remain to be explored further.

On May 23 2018, the LHCb Collaboration brought us a surprise with the observation of $\Xi_b(6227)^-$ in both the $\Lambda_b^0 K^-$ and $\Xi_b^0\pi^-$ invariant mass spectra \cite{Aaij:2018yqz}. As indicated in our former work~\cite{Chen:2018orb}, this newly discovered $\Xi_b(6227)^-$ may play a crucial role to construct the excited bottom baryon family. To some extent, the $\Xi_b(6227)^-$ becomes an important scaling point of establishing the whole excited bottom baryon states. In Ref.~\cite{Chen:2018orb}, $\Xi_b(6227)^-$ was suggested as a $P$-wave $\Xi_b^\prime$ baryon with $J^P=3/2^-$ or $5/2^-$. This conclusion was partly supported by the further theoretical work \cite{Aliev:2018lcs}.

Very recently, LHCb announced a new observation of the bottomed resonance structures, named the $\Sigma_b(6097)^{\pm}$, in the $\Lambda_b^0\pi^\pm$ final states from $pp$ collision \cite{Aaij:2018tnn}. Their resonance parameters were given as
\begin{eqnarray}
m_{\Sigma_b(6097)^-}&=&6098.0\pm1.7\pm0.5 \textrm{~MeV},\\
m_{\Sigma_b(6097)^+}&=&6095.8\pm1.7\pm0.4 \textrm{~MeV},\\
\Gamma_{\Sigma_b(6097)^-}&=&28.9\pm4.2\pm0.9 \textrm{~MeV},\\
\Gamma_{\Sigma_b(6097)^+}&=&31.0\pm5.5\pm0.7 \textrm{~MeV}.
\end{eqnarray}
For simplicity, hereafter we omit the charge signs of the discussed baryon states. First, we notice an interesting phenomenon, i.e., the mass difference of $\Sigma_b(6097)$ and $\Xi_b(6227)$ is around 130 MeV. The $\Sigma_b(6097)$ has the $bqq$ quark component ($q$ denotes $u$ and $d$ quark) while $\Xi_b(6227)$ has the $bsq$ quark component. Then the mass difference of $\Sigma_b(6097)$ and $\Xi_b(6227)$ is just equal to the mass difference of $u/d$ and $s$ constituent quarks.  Second, the observed decay modes exhibit the similarity of the decay behaviors for the $\Sigma_b(6097)$ and $\Xi_b(6227)$.\footnote{Because of the SU(3) flavor symmetry, the $\Lambda_b^0$ and $\Xi_b^{0,-}$ states are categorized into the antisymmetric antitriplet $\bar{3}_F$, which means they have the same quantum number assignment. The case of $\pi$ and $K$ is alike because they are the $1^1S_0$ states. According to the ordinary strong decay models, such as the $^3P_0$ model~\cite{Micu:1968mk,LeYaouanc:1972vsx,LeYaouanc:1988fx}, the decay amplitudes of the processes, $\Sigma_b(6097)\rightarrow\Lambda_b\pi$ and $\Xi_b(6227)\rightarrow\Lambda_bK/\Xi_b\pi$, have the same expression.} Two evidences lead us to ask whether the $\Sigma_b(6097)$ state is the nonstrange partner of $\Xi_b(6227)$ or not.

To answer this question, we adopt the same parameters given in our former work \cite{Chen:2018orb} as input and calculate the masses of excited nonstrange bottom baryons, i.e., $\Lambda_b$ and $\Sigma_b$ states. The obtained masses indeed support the conjecture of the $\Sigma_b(6097)$ above. Besides the proof from the mass spectrum analysis, the study of Okubo-Zweig-Iizuka (OZI)-allowed decays provides further support for this scenario. In the next section, we will illustrate the details.

With the success of decoding the inner structures of $\Sigma_b(6097)$ and $\Xi_b(6227)$, we have reason to believe that their $\Omega_b$ partner with the $ssb$ quark component must exist. For completeness, we continue to predict the mass positions and the corresponding two-body OZI-allowed decay behaviors of the 2$S$ and 1$P$ $\Omega_b$ states under the same theoretical framework. The results are valuable for experiment to further explore these unknown low-lying $\Omega_b$ baryons, which may be considered as the task of LHCb in the next stage. Surely, the research will provide a good opportunity to test our theoretical model.

In Ref.~\cite{Chen:2018orb}, we have emphasized the role of the $\Xi_b(6227)$ state for constructing the spectrum of the bottom baryon family. As the first excited bottom baryon state that was observed in the OZI-allowed decay channels, the observation of $\Xi_b(6227)$ can not only shed some light on the information of the mass spectra of the $P$-wave bottom baryons, but also provide some important clues for studying their decay properties. So we regard the $\Xi_b(6227)$ as the first scaling point when we try to build up the complete excited bottom baryon spectrum. Accordingly, the newly observed $\Sigma_b(6097)$ state could be the second scaling point which can help us to further understand the unknown excited bottom baryons.

The paper is organized as follows. After the introduction, we decode the newly reported $\Sigma_b(6097)$ as a $P$-wave state in Sec. \ref{sec2}, where the analysis of the mass spectrum and the investigation of strong decay can provide direct support to this scenario. In Sec. \ref{sec3}, the predicted masses and strong decays of the 2$S$ and 1$P$ $\Omega_b$ baryons are presented. Finally, the paper ends with the discussion and conclusion in Sec. \ref{sec4}.

\section{The $1P$ bottom baryon state assignment to the $\Sigma_b(6097)$ and its partner properties}\label{sec2}

To investigate the possible assignment of $\Sigma_b(6097)$, we first give a calculation of the $bqq$ baryon masses. The method, which was developed to successfully depict the mass spectrum of charm baryons in our previous papers~\cite{Chen:2016iyi,Chen:2017gnu}, will be employed here. We need to emphasize that this approach~\cite{Chen:2016iyi,Chen:2017gnu} was also applied to obtain the mass spectrum of the 1$S$, 2$S$, 3$S$, 1$P$, 2$P$, and 1$D$ $bsq$ baryons recently, which shows the newly announced $\Xi_b(6227)$ is a good candidate of the $P$-wave $\Xi_b^\prime$ baryon~\cite{Chen:2018orb}, where our spin-weighted average mass of the 1$P$ $\Xi_b$ states was supported by the recent work~\cite{Karliner:2018bms}. Additionally, the measured masses of 1$S$ $bsq$ states, $\Xi_b(5795)^-$, $\Xi^\prime_b(5935)^-$, and $\Xi^\ast_b(5955)^-$, can be reproduced well.

Under our framework, the $\lambda$-mode single heavy baryon system is simplified as a quasi-two-body system.
Then, in the heavy quark-light diquark picture, the following Schr\"odinger equation
\begin{equation}
\left(-\frac{\nabla^2}{2m_\mu}-\frac{4\alpha}{3r}+br+C+\frac{32\alpha\sigma^3}{9\sqrt{\pi}m_{di}m_b}\vec{\textrm{S}}_{di}\cdot\vec{\textrm{S}}_{b} \right)\psi_{nL} = E\psi_{nL}, \label{eq5}
\end{equation}
could be used to describe the dynamics between the light diquark and the heavy quark. When the spin-orbit and tensor interactions are treated as the leading-order perturbations, the masses of $\lambda$-mode heavy baryon states can be calculated. Interested readers can consult Refs.~\cite{Chen:2016iyi,Chen:2017gnu} for more details.

\begin{figure}[htbp]
\begin{center}
\includegraphics[width=8.6cm,keepaspectratio]{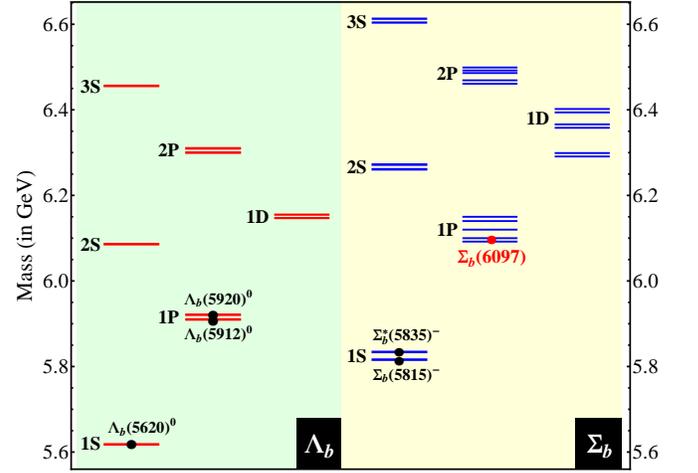}
\caption{The calculated masses for the $\Lambda_b$ and $\Sigma_b$ baryons and comparison with the experimental data. Here,  the observed $bqq$ baryons~\cite{Tanabashi:2018oca} and the newly reported $\Sigma_b(6097)$~\cite{Aaij:2018tnn}) are marked by the solid circles.}\label{Fig1}
\end{center}
\end{figure}

By taking the same parameters as that in Ref.~\cite{Chen:2018orb} as input and solving Eq.~(\ref{eq5}), we get the masses of $\Lambda_b$ and $\Sigma_b$ states, which are shown in Fig.~\ref{Fig1}. Here, for convenience, we list the values of the parameters in Table \ref{table1}. Obviously, the masses of $1S$ states, $\Lambda_b(5620)^0$, $\Sigma_b(5815)^-$, and $\Sigma^\ast_b(5835)^-$, in addition to the $P$-wave $\Lambda_b$ states, $\Lambda_b(5912)^0$ and $\Lambda_b(5920)^0$, have been well described by our method. Accordingly, the newly reported state, $\Sigma_b(6097)$, could be regarded as a 1$P$ state in our scheme. Indeed, the masses of $\Sigma_{b2}(3/2^-)$ and $\Sigma_{b2}(5/2^-)$ states are predicted to be 6094 and 6098 MeV, respectively,\footnote{We introduce the notation $\Sigma_{bJ_{di}}(J^P)$ to distinguish the five 1$P$ $\Sigma_b$ states, where $J_{di}$ is defined as the total angular momentum of the light diquark, i.e., $J_{di}\equiv\vec{S}_{di}+\vec{L}$.} which are in excellent agreement with the measurement of $\Sigma_b(6097)$.

\begin{table}[htbp]
\caption{The parameters for the bottom baryons adopted in our potential model. Here, the mass of the $b$ quark is taken as 4.96 GeV, and $m_{di}$ refers to the mass of a different diquark.}\label{table1}
\renewcommand\arraystretch{1.3}
\begin{tabular*}{86mm}{c@{\extracolsep{\fill}}ccccc}
\toprule[1pt]\toprule[1pt]
 Parameters       & $m_{di}$ (GeV)    & $\alpha$       & $b$  (GeV$^2$)   & $\sigma$ (GeV)    & $C$ (GeV)  \\
\toprule[1pt]
$\Lambda_b$       & 0.45              & 0.20           & 0.112            & $\cdots$          & 0.265  \\
$\Xi_b$           & 0.63              & 0.26           & 0.118            & $\cdots$          & 0.176 \\
$\Sigma_b$        & 0.66              & 0.22           & 0.116            & 1.20              & 0.185  \\
$\Xi^{\prime}_b$  & 0.78              & 0.22           & 0.116            & 1.20              & 0.152  \\
$\Omega_b$        & 0.91              & 0.26           & 0.120            & 1.07              & 0.120  \\
\bottomrule[1pt]\bottomrule[1pt]
\end{tabular*}
\end{table}

We may point out that another newly reported state, $\Xi_b(6227)$~\cite{Aaij:2018yqz}, has been suggested as a $P$-wave state with $J^P=3/2^-$ or $5/2^-$~\cite{Chen:2018orb}. In our scheme, the $\Sigma_b(6097)$ and $\Xi_b(6227)$ states may have the same $J^P$ assignments. This conclusion is partly supported by their mass gap. Specifically, the mass of $\Xi_b(6227)$ is about 130 MeV larger than the $\Sigma_b(6097)$, which is nearly equal to the mass difference between the $1S$ $\Xi_b^\prime$ and $\Sigma_b$ states,
\begin{eqnarray}
m_{\Xi^\prime_b(5935)^-}-m_{\Sigma_b(5815)^-}&\approx&119.5 \textrm{~MeV},\\
m_{\Xi^\ast_b(5955)^-}-m_{\Sigma^\ast_b(5835)^-}&\approx&120.2 \textrm{~MeV}.
\end{eqnarray}

Until now, many theoretical groups have paid attention to the mass spectrum of excited bottomed baryons. Among them, the nonrelativistic quark model~\cite{Kalman:1983an,Ghalenovi:2018fxh,Roberts:2007ni,Yoshida:2015tia}, the QCD-motivated relativistic quark model~\cite{Ebert:2011kk}, the QCD motivated hypercentral quark model~\cite{Thakkar:2016dna}, the relativized quark model with chromodynamics~\cite{Capstick:1986bm}, the Regge phenomenology~\cite{Wei:2016jyk}, the Faddeev methods~\cite{SilvestreBrac:1996bg,Valcarce:2008dr}, the relativistic flux tube model~\cite{Chen:2014nyo}, the QCD sum rule~\cite{Wang:2010it,Mao:2015gya}, and other method~\cite{Karliner:2018bms,Karliner:2015ema} were used. Especially, the masses of the $P$-wave $\Sigma_b$ states with $J^P=3/2^-$ and $J^P=5/2^-$ have been calculated in Refs. ~\cite{Kalman:1983an,Roberts:2007ni,Yoshida:2015tia,Ghalenovi:2018fxh,Capstick:1986bm,Ebert:2011kk,Thakkar:2016dna,Wei:2016jyk,SilvestreBrac:1996bg,Karliner:2015ema}, where the results were given in the range 6.08$-$6.18 GeV. Considering the intrinsic uncertainties of models, the results from these works do not contradict our suggested assignment of $\Sigma_b(6097)$. With the method of the QCD sum rule~\cite{Mao:2015gya}, the masses of $P$-wave $3/2^-$ and $5/2^-$ $\Sigma_b$ states were predicted to be $5.96\pm0.18$ and $5.98\pm0.18$ GeV, which also support the suggested assignment of $\Sigma_b(6097)$ above.

The mass spectrum analysis is only the first step to revealing the inner structure of $\Sigma_b(6097)$. In the following, the investigation of its strong decays will provide further information to distinguish the different $J^P$ quantum number assignment of $\Sigma_b(6097)$, which is a further step. Here, we employ the $^3P_0$ model to investigate the strong decay behaviors of these 1$P$ and 2$S$ $bqq$ baryon states.

The $^3P_0$ model has been extensively used to study the OZI-allowed strong decays of different kinds of hadrons. In Refs.~\cite{Chen:2016iyi,Chen:2017gnu}, we have studied the strong decays of these low-lying charmed baryons by the $^3P_0$ model. Recently, the $^3P_0$ model model was also used to calculate the decays of 1$P$ and 2$S$ $bsq$ baryon states~\cite{Chen:2018orb}. In the following, the strong decays of excited $\Lambda_b$ and $\Sigma_b$ baryon states will be given in the same framework. More details of the $^3P_0$ model and the input parameters in the calculation can be found in Refs.~\cite{Chen:2016iyi,Chen:2018orb}. We directly list the results of the strong decays of these discussed $bqq$ baryon states in Tables \ref{table2}-\ref{table3}, where the predicted masses corresponding to these bottom baryons are also attached.

\begin{table}[htbp]
\caption{The partial and total decay widths of the $1P$ $\Sigma_b$ states (in MeV). The forbidden decay channels are denoted by the symbol ``$\times$''. The measured decay width of $\Sigma_b(6097)^+$ is listed for comparison.} \label{table2}
\renewcommand\arraystretch{1.3}
\begin{tabular*}{86mm}{@{\extracolsep{\fill}}lccccc}
\toprule[1pt]\toprule[1pt]
Decay  &\multicolumn{2}{c}{$1/2^-$}  & \multicolumn{2}{c}{$3/2^-$}  & $5/2^-$  \\
\cline{2-3}\cline{4-5}\cline{6-6}
modes                        & $\Sigma_{b0}(6150)$   & $\Sigma_{b1}(6134)$ & $\Sigma_{b1}(6139)$ & $\Sigma_{b2}(6094)$ & $\Sigma_{b2}(6098)$   \\
\midrule[0.8pt]
 $\Lambda_b(5620)\pi$        &  10.4                 & $\times$            & $\times$            & 35.2                &  35.8    \\
 $\Sigma_b(5815)\pi$         & $\times$              & 83.8                & 4.5                 & 3.7                 &  1.8      \\
 $\Sigma^{*}_b(5835)\pi$     & $\times$              & 6.6                 & 90.8                & 2.6                 &  4.4        \\
 $\Lambda_b(5912)\pi$        &  5.9                  & 13.3                & 3.3                 & 0.7                 &  0.0   \\
 $\Lambda_b(5920)\pi$        &  10.9                 & 5.6                 & 15.8                & 0.1                 &  0.8     \\
 \midrule[0.8pt]
  Theory                     & 27.2                  &  109.3              & 114.4               & 42.3                &  42.8      \\
  Expt.~\cite{Aaij:2018yqz}  &                       &                     &                     & \multicolumn{2}{c}{$31.0\pm5.5\pm0.7$}        \\
\bottomrule[1pt]\bottomrule[1pt]
\end{tabular*}
\end{table}

Taking a glimpse of the results in Table \ref{table2}, one may conclude that the $\Sigma_b(6097)$ could be regarded as the $\Sigma_{b0}$ state, since the $\Lambda_b(5620)\pi$ is a main decay channel for this state and the total decay width of the $\Sigma_{b0}$ state here is 27.2 MeV, which is consistent with the experimental data of $\Sigma_b(6097)$.
However, the decay width of the $\Sigma_{b0}$ state is obtained by the predicted mass, i.e., 6150 MeV. When we take the measured mass of $\Sigma_b(6097)$ as input, the partial widths of the $\Lambda_b(5620)\pi$, $\Lambda_b(5912)\pi$, and $\Lambda_b(5920)\pi$ channels are predicted to be 2.1, 2.2, and 3.3 MeV, respectively.
Consequently, the corresponding total width is only 7.6 MeV, which is much smaller than the measured width of $\Sigma_b(6097)$. By this study, we may conclude that the $\Sigma_b(6097)$ seems not to be a $\Sigma_{b0}$ candidate.

Indeed, the $\Sigma_b(6097)$ is most likely a $P$-wave state with $J^P=3/2^-$ or $J^P=5/2^-$. As shown in Fig.~\ref{Fig1} and Table \ref{table2}, not only the predicted masses of $\Sigma_{b2}(3/2^-)$ and $\Sigma_{b2}(5/2^-)$ are in agreement with the measurement of the $\Sigma_b(6097)$, but also the predicted widths are comparable with the experimental result. The largest decay channel of $\Lambda_b(5620)\pi$ just shown in Table \ref{table2} also explains why the $\Sigma_b(6097)$ was first observed in this decay mode~\cite{Aaij:2018tnn}. The decay widths predicted by a constituent quark model~\cite{Wang:2017kfr} also support the suggested assignment to the $\Sigma_b(6097)$.

At present, we cannot distinguish the $J^P=3/2^-$ and $J^P=5/2^-$ quantum number assignments to the $\Sigma_b(6097)$ by comparing our theoretical result with the present experimental data. It is obvious that more accurate measurements are required in the near future, i.e.,  more decay modes, such as $\Sigma_b(5815)\pi$ and $\Sigma^{*}_b(5835)\pi$, are needed to be examined for these states. Especially, we suggest searching for the $\Sigma_{b1}(1/2^-)$ and $\Sigma_{b1}(3/2^-)$ in the decay channels of $\Sigma_b(5815)\pi$ and $\Sigma^{*}_b(5835)\pi$.

\begin{table}[htbp]
\caption{The partial and total decay widths of the $2S$ $\Lambda_b$ and $\Sigma_b$ states (in MeV). If the threshold of a decay channel lies above the excited state, we denote the channel by ``$-$''.} \label{table3}
\renewcommand\arraystretch{1.3}
\begin{tabular*}{85mm}{@{\extracolsep{\fill}}cccc}
\toprule[1pt]\toprule[1pt]
Decay modes                    & $\Lambda_b(6086)$  & $\Sigma_b(6261)$  & $\Sigma_b(6272)$ \\
\midrule[0.8pt]
 $\Lambda_b(5620)\pi$          & $\times$           & 26.6     & 25.8         \\
 $\Sigma_b(5815)\pi$           &   15.0             & 51.5     & 13.3        \\
 $\Sigma^\ast_b(5835)\pi$      &   20.6             & 23.1     & 60.0        \\
 $\Lambda_b(5912)\pi$          & $\times$           & 4.8      & 4.0          \\
 $\Lambda_b(5920)\pi$          & $\times$           & 8.5      & 9.0          \\
 $\Sigma_{b2}(6094)\pi$        & $-$                & 0.1      & 0.1          \\
 $\Sigma_{b2}(6098)\pi$        & $-$                & 0.1      & 0.2          \\
 $B N$                         & $-$                & 1.5      & 3.9            \\
 $B^\ast N$                    & $-$                & $-$      & 6.1            \\
 \midrule[0.8pt]
  Total                        & 35.6               & 116.2    & 122.4     \\
\bottomrule[1pt]\bottomrule[1pt]
\end{tabular*}
\end{table}

In this work, we also predict the existence of three $2S$ $bqq$ baryon states, which have the typical strong decay properties as listed in Table \ref{table3},
Our result indicates that the 2$S$ state, $\Lambda_b(6086)$, has two dominant decay channels of $\Sigma_b(5815)\pi$ and $\Sigma^\ast_b(5835)\pi$. The predicted partial decay width of $\Lambda_b(6086)\to \Sigma_b(5815)\pi$ is comparable with that of $\Lambda_b(6086)\to \Sigma^\ast_b(5835)\pi$. Decay widths of the remaining two 2$S$ bottom baryons, $\Sigma_b(6261)$  and $\Sigma_b(6272)$, are predicted to be about 120 MeV. The sizable
$\Lambda_b(5620)\pi$, $\Sigma_b(5815)\pi$, and $\Sigma^\ast_b(5835)\pi$ decay modes are ideal channels of searching for these missing 2$S$ $\Sigma_b$ baryons. The information given here is valuable for the LHCb to explore these unknown $2S$ $bqq$ baryon states.

\section{The predicted 1$P$ and 2$S$ $\Omega_b$ baryon states}\label{sec3}

The recent observations of $\Xi_b(6227)$ \cite{Aaij:2018yqz} and $\Sigma_b(6097)$ \cite{Aaij:2018tnn} enforce our confidence
to construct the whole high excited bottom baryon family. We have reason to expect that some excited $\Omega_b$ states can also be detected by LHCb in the near future.
Considering the present situation, theorists need to propose a new task for the experimentalist, and tell the experimental colleague how to find them.

Based on the study in Ref. \cite{Chen:2018orb} and in Sec. \ref{sec2} in this work, we continue to predict the 1$P$ and 2$S$ $\Omega_b$ bottom baryon states, which are as the partners of $\Xi_b(6227)$ \cite{Aaij:2018yqz} and $\Sigma_b(6097)$ \cite{Aaij:2018tnn}. Frankly speaking, the $\Xi_b(6227)$ \cite{Aaij:2018yqz} and $\Sigma_b(6097)$ \cite{Aaij:2018tnn} play the role of scaling points.

In this section, we also adopt the parameters in Table \ref{table1} to calculate the masses of excited $\Omega_b$ states. Our results are collected in Table~\ref{table4} with the results from other theoretical groups~\cite{Thakkar:2016dna,Ebert:2011kk,Roberts:2007ni,Yoshida:2015tia}. By the comparison in Table~\ref{table4},  we find that our results are comparable with these results obtained by other theoretical approaches. In addition, the masses of the 2$S$ $\Omega_b$ states obtained by the QCD sum rule~\cite{Agaev:2017jyt} are also comparable with the results in Table~\ref{table4}.

\begin{table}[htbp]
\caption{The predicted masses for the $\Omega_b$ baryons (in MeV). We collect the experimental values~\cite{Tanabashi:2018oca} and other theoretical results~\cite{Thakkar:2016dna,Ebert:2011kk,Roberts:2007ni,Yoshida:2015tia} for comparison. The states with the same $J^P$ but different masses are distinguished by the subscripts ``$h$'' and ``$l$''. The former one refers to the heavier state while the latter one denotes the lighter state.}\label{table4}
\renewcommand\arraystretch{1.3}
\begin{tabular*}{86mm}{@{\extracolsep{\fill}}ccccccc}
\toprule[1pt]\toprule[1pt]
 n$L~(J^P)$ & PDG~\cite{Tanabashi:2018oca} &   Our  & Ref.~\cite{Thakkar:2016dna} & Ref.~\cite{Ebert:2011kk} &  Ref.~\cite{Roberts:2007ni}  &  Ref.~\cite{Yoshida:2015tia}  \\
\midrule[0.8pt]
 $1S(1/2^+)$      & 6046.1      & 6045     & 6048    & 6064    & 6081    & 6076 \\
 $1S(3/2^+)$      &             & 6065     & 6086    & 6088    & 6102    & 6094 \\
 $2S(1/2^+)$      &             & 6483     & 6455    & 6450    & 6472    & 6517 \\
 $2S(3/2^+)$      &             & 6495     & 6481    & 6461    & 6478    & 6528 \\
 $1P(1/2^-)_h$    &             & 6361     & 6343    & 6339    & $-$     & 6340  \\
 $1P(1/2^-)_l$    &             & 6352     & 6338    & 6330    & 6388    & 6333 \\
 $1P(3/2^-)_h$    &             & 6363     & 6333    & 6340    & 6390    & 6336 \\
 $1P(3/2^-)_l$    &             & 6344     & 6328    & 6331    & 6304    & 6334 \\
 $1P(5/2^-)$      &             & 6349     & 6320    & 6334    & 6311    & 6345 \\
\bottomrule[1pt]\bottomrule[1pt]
\end{tabular*}
\end{table}

Furthermore, we present the decay behaviors of these predicted 2$S$ and 1$P$ $\Omega_b$ states by the $^3P_0$ model. Since the predicted masses of the five $1P$ $\Omega_b$ states are below the threshold of $\Xi_b^\prime(5935)K$ (see Table \ref{table4}), only the $\Xi_b(5795)K$ (denoted as $\Xi_bK$ in Table \ref{table5}) is allowed for the $P$-wave $\Omega_b$ states. The concrete results of their strong decays are given in Table \ref{table5}. The decays of $\Omega_{b1}(6352)$ and $\Omega_{b1}(6363)$ are strongly suppressed due to the constraint from the heavy quark symmetry. The widths of $\Omega_{b2}(6344)$ and $\Omega_{b2}(6349)$ are predicted as 2.3 and 2.7 MeV, respectively. So they are anticipated to be the narrow resonance structures in our scheme. A similar conclusion is also obtained in Ref.~\cite{Wang:2017kfr}. Another state, $\Omega_{b0}(6361)$, could also decay into $\Lambda_b(5620)K$. Its decay width is predicted about 33.4 MeV.

\begin{table}[htbp]
\caption{The decay widths of the $1P$ $\Omega_b$ states (in MeV).} \label{table5}
\renewcommand\arraystretch{1.3}
\begin{tabular*}{86mm}{@{\extracolsep{\fill}}lccccc}
\toprule[1pt]\toprule[1pt]
Decay  &\multicolumn{2}{c}{$1/2^-$}  & \multicolumn{2}{c}{$3/2^-$}  & $5/2^-$  \\
\cline{2-3}\cline{4-5}\cline{6-6}
modes                        & $\Omega_{b0}(6361)$   & $\Omega_{b1}(6352)$ & $\Omega_{b1}(6363)$ & $\Omega_{b2}(6344)$ & $\Omega_{b2}(6349)$   \\
\midrule[0.8pt]
 $\Xi_bK$          &  33.4                 & $\times$            & $\times$            & 2.3                &  2.7    \\
\bottomrule[1pt]\bottomrule[1pt]
\end{tabular*}
\end{table}

The obtained partial and total decay widths of two 2$S$ $\Omega_b$ states are listed in Table \ref{table6}. We can find that both states shall predominantly decay into $\Xi_b(5795)K$. The channels of $\Xi^\prime_b(5935)K$ and $\Xi^\ast_b(5955)K$ are also allowed. Their total decay widths are predicted as 15.5 and 16.1 MeV, respectively.

\begin{table}[htbp]
\caption{The predicted widths of 2$S$ $\Omega_b$ baryons (in MeV).}\label{table6}
\renewcommand\arraystretch{1.3}
\begin{tabular*}{86mm}{@{\extracolsep{\fill}}ccccc}
\toprule[1pt]\toprule[1pt]
 2$S$ $\Omega_b$ states  & $\Xi_b(5795)K$ &   $\Xi^\prime_b(5935)K$  & $\Xi^\ast_b(5955)K$ & Total  \\
\midrule[0.8pt]
 $\Omega_b(6483)$      & 12.6           &      2.4                 &         0.5         & 15.5    \\
 $\Omega_b(6495)$      & 13.2           &      0.8                 &         2.1         & 16.1    \\
\bottomrule[1pt]\bottomrule[1pt]
\end{tabular*}
\end{table}

Until now, only the $\Omega_b(6046)$ has been reported by experiment~\cite{Tanabashi:2018oca}, which is a $1S$ state in the $\Omega_b$ baryon family. When the above information was given, the experimentalist should have the ambition of searching for these missing $\Omega_b$ bottom baryons. Although it is full of the opportunity, it is obvious that there also exist challenge, especially for these narrow states. Additionally, it is not the end of whole story for theorist. Further theoretical efforts from different points of view will be helpful to explore these $\Omega_b$ bottom baryons.

\section{Discussions and conclusions}\label{sec4}

The newly observation of $\Sigma_b(6097)$ \cite{Aaij:2018tnn} inspired our interest in decoding its inner structure. We realize that
the $\Sigma_b(6097)$ can be the second scaling point when constructing the whole excited bottom baryon family.
Indeed the phenomenological analysis presented in this work shows that the $\Sigma_b(6097)$ is a $1P$ bottom baryon candidate with either $J^P=3/2^-$ or $J^P=5/2^-$. This fact indicates that the $\Sigma_b(6097)$ state is the nonstrange partner of $\Xi_b(6227)$ \cite{Aaij:2018yqz}, since both $\Sigma_b(6097)$ and $\Xi_b(6227)$ have the similar decay behavior and spin-parity quantum number just shown in the present work and in Ref. \cite{Chen:2018orb}. It is a crucial progress on establishing the high excited states of bottom baryon family.

When explaining the $\Sigma_b(6097)$, we predict its $1P$ and $2S$ bottom baryon partners, which are still missing in experiment.
Our study has not only given their mass positions, but also illustrated the key decay channels reflected by the obtained partial decay widths.

In fact, the success of explaining $\Sigma_b(6097)$ and $\Xi_b(6227)$ shows the reasonability of the quasi-two-body treatment for the heavy baryons. Thus, we employ the theoretical models to investigate the $\Omega_b$ system, where we further predict the mass spectra of the $2S$ and $1P$ $\Omega_b$ states and the corresponding strong decay behaviors. The experimental search for these $\Omega_b$ states will be an interesting research issue, especially at the LHCb.

\begin{figure}[htbp]
\begin{center}
\includegraphics[width=8.5cm,keepaspectratio]{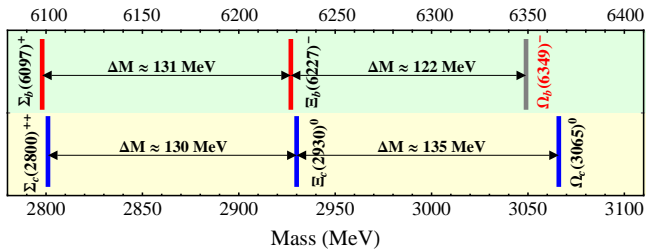}
\caption{The mass gaps of the discovered heavy baryons, $\Sigma_c(2800)^{++}$, $\Xi_c(2930)^0$, $\Omega_c(3065)^0$, $\Sigma_b(6097)^+$, $\Xi_b(6227)^-$, and the predicted $\Omega_b(6349)^-$ state.}\label{Fig2}
\end{center}
\end{figure}

Before ending the paper, we present a comparison of some charm and bottom baryons in Fig.~\ref{Fig2}.
It is interesting to point out that three charm baryons, i.e., $\Sigma_c(2800)$~\cite{Mizuk:2004yu}, $\Xi_c(2930)$~\cite{Aubert:2007eb,Li:2017uvv,Li:2018fmq}, and $\Omega_c(3065)$~\cite{Aaij:2017nav,Yelton:2017qxg}, have been assigned as the 1$P$ candidates with $J^P=3/2^-$ or $5/2^-$ by our previous efforts~\cite{Chen:2016iyi,Chen:2017gnu}. The newly reported state, $\Xi_b(6227)$, also favors the same $J^P$ assignment~\cite{Chen:2018orb}. To compare their masses clearly, we summarize them in addition to $\Sigma_b(6097)$ in Fig.~\ref{Fig2}. Obviously, their mass gaps are about 130 MeV. This result could be explained as the mass difference of the $u/d$ and $s$ constituent quarks. Then, we may predict an unknown $\Omega_b$ state which is the $bss$ partner of $\Sigma_c(2800)$, $\Xi_c(2930)$, $\Omega_c(3065)$, $\Sigma_b(6097)$, and $\Xi_b(6227)$. With the predicted mass by our method, we denote this state as $\Omega_b(6349)$ in Fig.~\ref{Fig2}. Therefore, finding the $\Omega_b(6349)$ state in the future experiment through analyzing the $\Xi_b(5795)K$ invariant mass spectrum will be a key point of testing the scenario proposed by our theoretical models.

With the discovery of the new states, $\Sigma_b(6097)$ and $\Xi_b(6227)$, LHCb has shown its capability in accumulating the data sample of the excited bottom baryon resonances. With the coming LHCb Upgrade I in 2020~\cite{Bediaga:2018lhg}, more and more excited $b$-flavor baryons are expected to be found. It is obvious that it is time for the LHCb.

\section*{Acknowledgement}

B.C. thanks Si-Qiang Luo for checking the result of strong decays. This project is supported by the National Natural Science Foundation of China under Grant No. 11305003, No. 11222547, No. 11175073, and No. 11647301. X.L. is also supported by the National  Program for Support of Top-notch Young Professionals.


\end{document}